\documentclass{iaus}
\usepackage{graphicx}

\newcommand{\hii}{{H{\scriptsize II} }}
\newcommand{\nhthree}{\mbox{NH$_3$}}
\newcommand{\nhone}{\mbox{NH$_3$(1,1)}}

\newcommand{\nhfour}{\mbox{NH$_3$(4,4)}}
\newcommand{\nhfive}{\mbox{NH$_3$(5,5)}}
\newcommand{\kms}{\mbox{kms$^{-1}$}}
\newcommand{\methanol}{\mbox{CH$_3$OH}}
\newcommand{\arcsec}{^{\prime\prime}}
\newcommand{\arcmin}{^{\prime}}

\title[The molecular environment associated with Class II methanol masers]
{The molecular environment of massive star forming cores associated
with Class II methanol maser emission}

\author[S. N. Longmore et al.]  
			 {S.~N.~Longmore$^{1,2}$\thanks{E-mail:snl@phys.unsw.edu.au},
			 M.~G.~Burton$^1$, P.~J.~Barnes$^{3}$,
			 T.~Wong$^{1,2,5}$, C.~R.~Purcell$^{1,4}$,
			 J.~Ott$^{2,6}$}

\affiliation{$^1$School of Physics, University of New South Wales, Kensington, NSW 2052, Sydney, Australia\\[\affilskip]
$^2$Australia Telescope National Facility, CSIRO, PO Box 76, Epping, NSW 1710, Australia\\[\affilskip]
$^3$School of Physics A28, University of Sydney, NSW 2006, Australia\\[\affilskip]
$^4$University of Manchester, Jodrell Bank Observatory, Macclesfield, Cheshire SK11 9DL, UK\\[\affilskip]
$^5$Department of Astronomy, University of Illinois, Urbana IL 61801, USA\\[\affilskip]
$^6$National Radio Astronomy Observatory, 520 Edgemont Road, Charlottesville, VA 22903, USA\\}

\pubyear{2007}
\volume{242}  
\pagerange{119--126}
\date{?? and in revised form ??}
\jname{Astrophysical masers and their environments}
\editors{A.C. Editor, B.D. Editor \& C.E. Editor, eds.}
\begin{document}

\maketitle

\begin{abstract}
Methanol maser emission has proven to be an excellent signpost of
regions undergoing massive star formation (MSF). To investigate their
role as an evolutionary tracer, we have recently completed a large
observing program with the ATCA to derive the dynamical and physical
properties of molecular/ionised gas towards a sample of MSF regions
traced by 6.7\,GHz methanol maser emission. We find that the molecular
gas in many of these regions breaks up into multiple sub-clumps which
we separate into groups based on their association with/without
methanol maser and cm continuum emission. The temperature and dynamic
state of the molecular gas is markedly different between the
groups. Based on these differences, we attempt to assess the
evolutionary state of the cores in the groups and thus investigate the
role of class II methanol masers as a tracer of MSF.

\end{abstract}

\firstsection 

\section{Introduction}

In terms of luminosity, energetics and chemical enrichment, massive
stars exert a disproportionate influence compared to their number on
Galactic evolution.  However, the collective effects of their rarity,
short formation timescales and heavy obscuration due to dust, make it
difficult to find large samples of massive young sources at well
constrained evolutionary stages needed to develop an understanding of
their formation mechanism. The 5$_1$$\, \rightarrow \,$6$_0$ A$^+$
class II methanol (\methanol) maser transition at 6.7\,GHz is one of
the most readily observable signposts of MSF
[\cite{menten1991}].

The specific conditions required for the masers to exist makes them a
powerful probe of the region's evolutionary stage. While masers probe
spatial scales much smaller than their natal cores, the numerous
feedback processes from newly formed stars [e.g.  turbulence injection
from jets/outflows, ionisation (stars M$>$8M$_\odot$) and heating from
radiation etc.]  must significantly alter the physical conditions of
the surrounding region. We have completed an observing program with
the Australia Telescope Compact Array (ATCA) to derive properties of
molecular and ionised gas towards MSF regions traced by 6.7\,GHz
methanol maser emission [\cite{longmore2007}]. In this contribution,
we use these results to investigate the use of class II methanol
masers as a diagnostic of the evolutionary stage of MSF.

\section{Observations}
Observations of $\nhone$, (2,2), (4,4) \& (5,5) and 24\,GHz continuum
were carried out using the ATCA towards 21 MSF regions traced by
6.7\,GHz methanol maser emission [selected from a similar sample to
Hill \etal\ (this volume)]. The H168 [$\nhone$/(2,2)] and H214
[$\nhfour$/(5,5)] antenna configurations with both East-West and
North-South baselines, were used to allow for snapshot
imaging. Primary and characteristic synthesised beam sizes were
$\sim$2.$\arcmin$2 and $\sim$$8 - 11 \arcsec$ respectively. Each
source was observed for 4$\times$15 minute cuts in each transition
separated over 8 hours to ensure the best possible sampling of the
$uv$-plane. The data were reduced using the MIRIAD \cite[(see Sault
\etal\ 1995)]{sault1995} package. Characteristic spectra were
extracted at every transition for each core at the position of the
peak $\nhone$ emission, baseline subtracted and fit using the
\emph{gauss} and \emph{nh3(1,1)} methods in CLASS (see
http://www.iram.fr/IRAMFR/GILDAS/). Continuum source fluxes and
angular sizes were calculated in both the image domain and directly
from the $uv$ data.

\section{Core separation}
\label{sec:core_separation}
$\nhthree$ detections within each region were separated into
individual cores if offset by more than a synthesised beam width
spatially, or more than the FWHM in velocity if at the same sky
position. The same criteria were used to determine whether the
$\nhthree$, continuum and methanol maser emission in each of the
regions were associated. In all but three cases, these criteria were
sufficient to both unambiguously separate cores and determine their
association with continuum and maser emission. We find 41 $\nhone$
cores (of which 3 are in absorption and 2 are separated in velocity)
and 14 24\,GHz continuum cores. Observationally the cores fall in to 4
groups: $\nhthree$ only (Group 1); $\nhthree$~+~methanol maser (Group
2); $\nhthree$~+~methanol maser~+~24\,GHz continuum (Group 3);
$\nhthree$~+~24\,GHz continuum (Group 4). The cores were distributed
with 16, 16, 6 and 2 cores in Groups 1 to 4, respectively. Based on
this grouping, most of the $\nhone$ cores are coincident with methanol
maser emission (Groups 2 \& 3), but there are a substantial fraction
of $\nhthree$ cores with neither 24\,GHz continuum nor maser emission
(Group 1).

Having separated the cores into these groups, we then considered
observational biases and selection effects which may affect their
distribution. The biggest potential hindrance was the difference in
linear resolution and sensitivity caused by the factor of $\sim$5
variation in distance to the regions. Despite this, the $\nhthree$,
continuum and methanol maser observations have the same sensitivity
limit towards all the regions: therefore, the \emph{relative} flux
densities of these tracers in a given region are directly
comparable. In addition, no correlation was found between a region's
distance and the number of cores toward the region or their
association with the different tracers. From this we conclude the
distance variation does not affect the distribution of cores into
separate groups. However, it should be remembered that any conclusions
drawn about the cores are limited by the observational parameters used
to define the groups.

\section{Deriving physical properties}

Properties of the molecular gas in each of the cores were derived from
the $\nhthree$ observations. The core size was calculated from the
extent of the integrated $\nhone$ emission after deconvolving the
synthesised beam response. The dynamical state of the molecular gas
was derived from the line profiles of the high spectral resolution
(0.197\,\kms) $\nhone$ observations after deconvolving the
instrumental response. Preliminary gas kinetic temperatures were
calculated by fitting the measured column densities of the multiple
$\nhthree$ transitions to the LVG models described
in~\cite{ott2005}. Finally, properties of the ionised gas were derived
from the 24\,GHz continuum emission
following~\cite{mezger_henderson1967}, assuming it was spherically
symmetric and optically thin at an electron temperature of 10$^4$\,K.

\section{Results}
In general the core properties are comparable to those derived from
similar surveys towards young MSF regions. Below we outline
differences, in particular between the cores in the different groups
described in $\S$\ref{sec:core_separation}.

\subsection{Molecular Gas}
The measured $\nhone$ linewidth varies significantly between the
groups, increasing from 1.43, 2.43, 3.00\,$\kms$ for Groups 1 to 3
respectively. This shows the $\nhthree$-only cores are more quiescent
than those with methanol maser emission.

The~\nhone~spectra of some cores deviate significantly from the
predicted symmetric inner and outer satellite brightness temperature
ratios. These line profile asymmetries are often seen toward star
forming cores and are understood to be caused by selective radiative
trapping due to multiple~\nhone~sub-clumps within the beam
[see~\cite{stutzki_winnewisser1985b} and references therein]. The
\nhthree-only cores (Group 1) have by far the strongest asymmetries.

$\nhfour$ emission is detected toward the peak of 13 $\nhone$ cores
and 11 of these also have coincident $\nhfive$ emission. The higher
spatial resolution of the $\nhfour$ and (5,5) images compared to the
$\nhone$ observations (8$\arcsec$ vs 11$\arcsec$) provides a stronger
constraint to the criteria outlined in $\S$\ref{sec:core_separation}
as to whether this emission is associated with either methanol or
continuum emission. In every case, the $\nhfour$ and (5,5) emission is
unresolved, within a synthesised beam width of the methanol maser
emission spatially and within the FWHM in velocity. This shows the
methanol masers form at the warmest parts of the core. Significantly,
no $\nhfour$ or (5,5) emission is detected toward $\nhthree$-only
sources.

\begin{figure}
    \begin{center}
      	  \includegraphics[width=5cm, angle=-90, trim=0 0 -5 0]{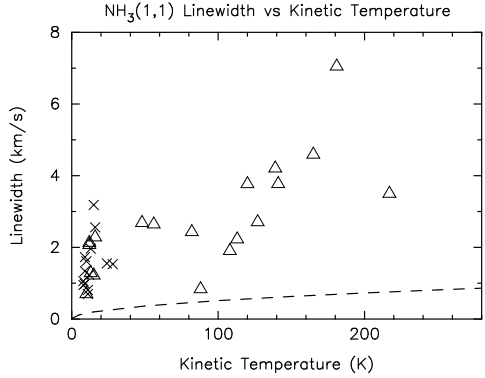}
    \end{center}
    \caption{$\nhone$ linewidth vs gas kinetic temperature. Cores with
    $\nhthree$ only (Group 1) are shown as crosses while those with
    $\nhthree$ and methanol maser emission (Groups 2 and 3) are shown
    as triangles. The dashed line shows the expected linewidth due to
    purely thermal motions.}
	\label{fig:temp}
\end{figure}

As shown in Figure~\ref{fig:temp}, cores with $\nhthree$ and methanol
maser emission (Groups 2 and 3) are generally significantly warmer
than those with only $\nhthree$ emission (Group 1). However, there are
also a small number of cores with methanol maser emission that have
very cool temperatures and quiescent gas, similar to the
$\nhthree$-only cores. Modelling shows pumping of 6.7\,GHz methanol
masers requires local temperatures sufficient to evaporate methanol
from the dust grains (T$\gtrsim$90K) and a highly luminous source of
IR photons [\cite{cragg2005}] i.e.  an internal powering source. It is
therefore plausible that the cold, quiescent sources with methanol
maser emission are cores in which the feedback from the powering
sources have not had time to significantly alter the larger scale
properties of the gas in the cores.

\subsection{Ionised Gas}
Of the 14 continuum cores detected at 24\,GHz, 10 are within two
synthesised beams of the 6.7GHz methanol maser emission. This is
contrary to the results of \cite{walsh1998}, who found the masers
generally unassociated with 8\,GHz continuum emission. However, six of
the 24\,GHz continuum sources found at the site of the methanol maser
emission have no 8\,GHz counterparts. A possible explanation for this
may be that the continuum emission is optically thick rather than
optically thin between 8 and 24\,GHz and hence has a flux density
proportional to $\nu^2$ rather than $\nu^{-0.1}$. The seemingly low
emission measures derived for the 24\,GHz continuum are unreliable due
to the large beam size of the observations. Alternatively, the 24\,GHz
continuum sources may have been too extended and resolved-out by the
larger array configuration used at 8\,GHz by \cite{walsh1998}. Further
high spatial resolution observations at $\nu \geq$ 24\,GHz are
required to derive reliable emission measures and spectral indexes to
unambiguously differentiate between the two explanations.

\section{Towards an evolutionary sequence}

From the previous analysis, the core properties are seen to vary
depending on their association with methanol maser and continuum
emission. Making the reasonable assumption that cores heat up and
becomes less quiescent with age, we now investigate what these
physical conditions tell us about their evolutionary state.

As the core sizes are similar, the measured linewidths can be reliably
used to indicate how quiescent the gas is, without worrying about its
dependence on the core size [\cite{larson1981}]. It then becomes
obvious that the isolated $\nhthree$ cores (Group 1) contain the most
quiescent gas. However, from the linewidths alone it is not clear if
these cores will eventually form stars or if they are a transitory
phenomenon. The fact that a large number of these Group 1 cores
contain many dense sub-clumps (as evidenced by the $\nhone$
asymmetries) suggests the former is likely for at least these
cores. The linewidth of sources with methanol maser emission (Groups 2
and 3) are significantly larger, and hence contain less quiescent gas,
than those of Group 1. The larger linewidths combined with generally
higher temperatures, suggests cores in Groups 2 and 3 are more evolved
than in Group 1.

The detection of continuum emission suggests a massive star is already
ionising the gas. With the current observations, the properties of the
continuum sources are not well enough constrained to further separate
their evolutionary stages. However, as all (with one possible
exception) of the continuum sources only detected at 24GHz are
associated with dense molecular gas and masers, this would suggest
they are younger than those detected at both 8 + 24\,GHz, despite
their seemingly small emission measures. In this scenario, the cores
with only 8 + 24GHz continuum and no $\nhthree$ emission, may be
sufficiently advanced for the UC\hii region to have destroyed its
natal molecular material.

From this evidence, the cores in the different groups do appear to be
at different evolutionary stages, going from most quiescent to most
evolved according to the group number.

\section{Conclusions}
What then, can we conclude about the role of methanol masers as a
signpost of MSF? The observations show that
6.7\,GHz methanol masers:
\begin{itemize}
  \item are found at the warmest location within each core.
  \item generally highlight significantly warmer regions with less
  quiescent gas (i.e. more evolved sources) than those with only
  $\nhthree$ emission.
  \item may also highlight regions in which the internal pumping
  source is sufficiently young that it has not yet detectably altered
  the large scale core properties.
\end{itemize}

Methanol masers therefore trace regions at stages shortly after a
suitable powering source has formed right through to relatively
evolved UC\hii~regions. While remaining a good general tracer of young
MSF regions, the presence of a methanol maser does
not single out any particular intermediate evolutionary stage.

Finally, these data confirm and strengthen the results of Hill \etal\
(this volume), that the youngest MSF regions appear to be molecular
cores with no detectable methanol maser emission.

\section{Acknowledgements}
SNL is supported by a scholarship from the School of Physics at
UNSW. We thank Andrew Walsh for comments on the manuscript. We also
thank the Australian Research Council for funding support. The
Australia Telescope is funded by the Commonwealth of Australia for
operation as a National Facility managed by CSIRO.

\end{document}